\documentclass[preprint,journal]{vgtc}            

\onlineid{1354}

\vgtccategory{Research}

\vgtcpapertype{please specify}

\title{Hold Tight: Identifying Behavioral Patterns During Prolonged Work in VR through Video Analysis}

\author{
 Verena Biener$^{1}$\thanks{e-mail: verena.biener@hs-coburg.de}
 \and Forouzan Farzinnejad$^{1}$
\and Rinaldo Schuster$^{1}$
\and Seyedmasih Tabaei$^{1}$
\and Leon Lindlein$^{1}$
\and Jinghui Hu$^{2}$
\and Negar Nouri$^{1}$
\and John J. Dudley$^{2}$
\and Per Ola Kristensson$^{2}$
\and Jörg Müller$^{3}$
\and Jens Grubert$^{1}$\thanks{e-mail: jens.grubert@hs-coburg.de}
}
\affiliation{\scriptsize $^{1}$Coburg University of Applied Sciences and Arts, Germany\\  $^{2}$University of Cambridge, United Kingdom \\  $^{3}$University of Bayreuth, Germany}

\usepackage{soul} 
\usepackage{url}
\usepackage{balance}

\newif\ifhighlightchanges

\newcommand{\newmarker}[1]{%
\ifhighlightchanges
\textcolor{blue}{#1}%
\else
#1%
\fi
}

\newcommand{\deletemarker}[1]{%
\ifhighlightchanges
\textcolor{red}{\st{#1}}%
\else
\fi
}

\highlightchangesfalse

\abstract{%
VR devices have recently been actively promoted as tools for knowledge workers and prior work has demonstrated that VR can support some knowledge worker tasks. 
However, only a few studies have explored the effects of prolonged use of VR 
\deletemarker{.While a prior study on working in VR for one week focused on reporting performance measures and subjective feedback, a nuanced understanding of participants' behavior in VR and how it evolves over time is still missing.}
\newmarker{such as a study observing 16 participants working in VR and a physical environment for one work-week each and reporting mainly on subjective feedback.   
As a nuanced understanding of participants' behavior in VR and how it evolves over time is still missing,} we report on the results from an analysis of 559 hours of video material \deletemarker{in which participants were working in VR for an entire workweek.}\newmarker{obtained in this prior study.}
Among other findings, we report that (1) the
frequency of actions related to adjusting the headset reduced by 46\% \deletemarker{and the duration by 50\% over the five days; (2)} \newmarker{and} the frequency of actions related to supporting the headset reduced by 42\% over the five days; (2) the HMD was removed 31\% less frequently over the five days but for 41\% longer periods; (3) wearing an HMD is disruptive to normal patterns of eating and drinking, but not to social interactions, such as talking.
The combined findings in this work demonstrate the value of long-term studies of deployed VR systems and can be used to inform the design of better, more ergonomic VR systems as tools for knowledge workers.
}


\teaser{
  \centering
  \includegraphics[width=\linewidth]{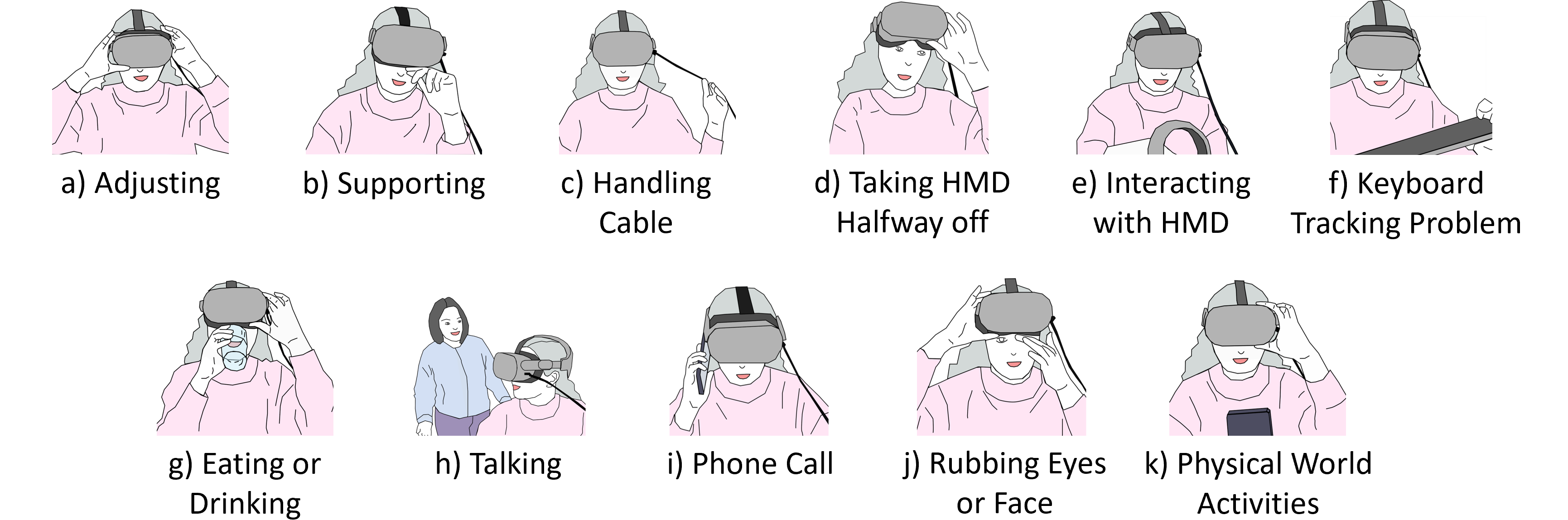}
  \caption{%
  	Categories of actions observed in the video-analysis: (a) adjusting the HMD; (b) supporting the HMD; (c) handling the HMD's power cable; (d) taking the HMD halfway off, by lifting it up a  bit; (e) interacting wit the HMD through controllers or gestures; (f) encountering keyboard tracking problems; (g) eating or drinking something; (h) talking to another physically present person; (i) having a phone call; (j) rubbing eyes or face; (k) doing physical world activities such as using a phone or reading from paper.%
  }
  \label{fig:teaser}
}

\graphicspath{{figs/}{figures/}{pictures/}{images/}{./}} 

\usepackage{tabu}                      
\usepackage{booktabs}                  
\usepackage{lipsum}                    
\usepackage{mwe}                       

\usepackage{mathptmx}                  

\begin{document}


\firstsection{Introduction}

\maketitle

Virtual Reality (VR) has already gained popularity in the entertainment domain, but it has also been explored in recent years as a tool for knowledge work ~\cite{biener2020breaking,ruvimova2020transport,biener2022povrpoint}.
VR can provide various advantages for improving work experiences, such as enhancing interactivity~\cite{biener2022povrpoint}, adapting work-environments~\cite{ruvimova2020transport} or utilizing large virtual displays~\cite{biener2020breaking, mcgill2020expanding}.
Still, the prospect of wearing current-generation VR headsets for a prolonged period of time, such as a full workday or even a whole workweek, could be off-putting due to the current state of VR systems.
When using VR applications intended for entertainment this may be less of an issue as users might be distracted from the less-than-optimal hardware by an engaging virtual experience. However, this might not be the case for knowledge workers who need to wear HMDs for prolonged periods of time.
This motivates research with the objective of understanding how knowledge workers respond to the prolonged use of VR.

To this end, Biener et al.~\cite{biener2022quantifying} conducted a study in which participants completed a full workweek in VR\deletemarker{while performing their usual work tasks.
This was compared to a week in which the same participants worked} \newmarker{and compared it to a week} in a comparable physical setup.
\deletemarker{Biener et al.}
\newmarker{They} found that the VR condition resulted in significantly worse ratings for measures of task load, frustration, negative affect, anxiety, eye strain, system usability, flow, productivity, well-being and simulator sickness.
Nevertheless, some of the reported measures improved slightly in the course of the five days.
While conducting this study, Biener et al.~\cite{biener2022quantifying} generated a dataset of over 1,400 hours of video material, capturing the participants' \deletemarker{faces and parts of their upper body, and therefore their} behavior in both conditions throughout the duration of both workweeks. However, this data was not analyzed in the original paper.
We have obtained access to this dataset of the original study and present an extensive analysis of the user behaviors exhibited in this rich video data.

\deletemarker{In this paper,} We report on the behavior of 16 participants as they respond to the experience of working in VR over a workweek, by sampling the first, third, and last day.\deletemarker{of their workweek.}
Where possible, we also compare this behavior with the observed behavior of participants in the non-VR condition. This resulted in a total of 559 hours of analyzed video data.
To our knowledge, this is the first paper reporting on such a comprehensive video study of VR work and it provides substantial\newmarker{ and detailed} insights into the behavior of users in a VR work setting\deletemarker{,}\newmarker{. It is} covering much longer time periods in comparison to typical VR studies\newmarker{, and, therefore, also allows us to describe how participants' behavior changes over time}.

\deletemarker{Lacking an automated process for annotating the 559 hours of video, six annotators manually reviewed and coded participants' behaviors observed in the video data.}
Common behaviors \newmarker{that we observed} include: adjusting\deletemarker{the headset}, supporting\deletemarker{the headset}, \newmarker{or} removing the headset, standing, eating and drinking, using a phone, or interacting with other people.
\deletemarker{We report on the extensive analysis of these behaviors in the remainder of the paper, 
but }Key insights \deletemarker{obtained} are: \newmarker{over the five days }(1) the frequency of actions related to adjusting the headset was reduced by 46\% and the duration by 50\%\deletemarker{over the five days}; (2) the frequency of actions related to supporting the headset was reduced by 42\%\deletemarker{over the five days}; (3) the HMD was removed 31\% less frequently\deletemarker{over the five days} but for 41\% longer periods; (4) wearing an HMD is disruptive to normal patterns of eating and drinking, but not to social interactions, such as talking.


These observations reveal a pattern of accommodation, adaptation, and appropriation \deletemarker{of what constitutes to workers}\newmarker{to} a novel working setup.
Viewed in concert with the findings of Biener et al.~\cite{biener2022quantifying}, however, these adaptations were insufficient to bridge the gap between the VR and physical work setups.
Our findings thereby provide evidence and motivation for a broader consideration of usability and ergonomic issues encountered by wearers of HMDs under extended use, which so far has been insufficiently explored in the literature.
The insights presented in this paper can thus inform both: (1) the physical design of HMDs to improve comfort and simplify adjustment; and (2) the development of ergonomic guidance for workers tasked to wear an HMD for any extended period of time.

\section{Related Work}
VR has several possible advantages and challenges as a tool for knowledge work which are discussed in Section \ref{sec:knowledgeWorkInVR}.
However, the studies investigating the benefits of VR are usually quite short. Therefore, Section \ref{sec:prolongedStudies} reviews previous work about using VR for longer periods of time. Then, Section \ref{sec:videoAnalysisStudies} reports on prior studies using video-analysis to uncover the behavior of participants.

\subsection{Knowledge Work in VR}
\label{sec:knowledgeWorkInVR}
For more than a decade VR has been used in industry to support work in various contexts \cite{berg2017industry}.
Recent VR HMDs have been promoted specifically as a tool for knowledge workers (e.g.~\cite{ofek2020towards}) and previous studies have also shown that VR has the potential to support knowledge work.
Biener et al.~\cite{biener2020breaking, biener2022povrpoint} have shown that multimodal interaction, combining touch and eye-tracking can be used to efficiently navigate multiple screens and that three-dimensional visualizations can facilitate tasks involving multiple layers of information.
In combination with a spatially tracked pen, VR has been shown to facilitate certain tasks in spreadsheet applications \cite{gesslein2020pen}, or when authoring presentations \cite{biener2022povrpoint}.
Other work has investigated how to use a 2D mouse to interact with 3D content in VR \cite{zhou2022depth}. 
Using eye-gaze and blink, Meng et al.~\cite{meng2022anexploration} explored techniques for hands-free text-selection in VR, and Lee et al.~\cite{lee2022vrdoc} proposed a technique that uses gaze to make reading in VR more efficient and less demanding.

In addition to novel interaction techniques, VR makes it possible to have any number of displays which could be especially helpful in a mobile context where the screen space of conventional devices is limited. 
However, after comparing physical and virtual monitors in AR, Pavanatto et al.~\cite{pavanatto2021we} suggested to use a mix of both, due to the current limitations of AR devices such as low resolution.
McGill et al.~\cite{mcgill2020expanding} suggested to manipulate the virtual display position using the users gaze-direction \deletemarker{to allow them} to use a large display space with less head movement.
\deletemarker{Virtual displays are especially beneficial in mobile settings with limited space, yet}
\newmarker{Even though virtual displays can be beneficial in settings with limited space,} Ng et al.~\cite{ng2021passenger} found that passengers in an airplane preferred to limit virtual displays to their personal seating area.
Similarly, Medeiros et al.~\cite{medeiros2022fromshielding} found that users avoid placing displays at the location of other passengers in a public transportation scenario, but that they use them to shield themselves from others.

VR can also address privacy issues when used in a public space, for example by randomizing the layout of a physical keyboard when entering a password \cite{schneider2019reconviguration}, or by the fact that only the person wearing the HMD can see the content on the virtual screens~\cite{grubert2018office}.

In addition, VR can be used to reduce distractions and stress.
Ruvimova et al.~\cite{ruvimova2020transport} reported that \deletemarker{using a VR office with} a virtual beach environment when sitting in an open office environment \deletemarker{reduced distractions and induced}\newmarker{can reduce distractions and induce} flow.
Similarly, Lee et al.~\cite{lee2019partitioning} showed that using AR to add visual separators in an open office reduces distractions and allows to easily personalize the work environment.
Thoondee et al.~\cite{thoondee2017usingvirtual} reported that participants who were experiencing a VR relaxation environment were more relaxed.
It has also been indicated that experiencing nature in VR can reduce stress better than watching a video on a regular display~\cite{pretsch2021improving}, and that interactive nature environments in VR have more positive effects on stress than passive VR experiences~\cite{valtchanov2010restorative}.
However, even though users prefer natural environments, they perform better in familiar work environments~\cite{li2021rear}.

This prior work demonstrates a range of positive aspects of working in VR, yet the results have been gathered through short-term studies. In addition, a review of previous work about office-like tasks in VR 
\cite{souchet2022narrative} found that VR could induce increased visual fatigue, muscle fatigue, acute stress and mental overload.
Therefore, further research is needed on the effects of working in VR, especially for extended periods of time.

VR can reduce distractions \cite{ruvimova2020transport} or help users relax \cite{thoondee2017usingvirtual}, by shielding them from the physical world.
On the other hand, there are also 
\deletemarker{objects or other people in the} 
\newmarker{aspects of the}
physical surroundings that the user should be aware of while working in VR.
To overcome this problem, McGill et al.~\cite{mcgill2015dose} suggested to integrate relevant parts of the physical environment into the virtual, so that users are aware of other people or relevant objects, such as a nearby cup of tea on the desk. To this end, Wang et al.~\cite{wang2022realitylens} present a customizable physical world view and Hartmann et al.~\cite{hartmann2019realitycheck} include real-time 3D reconstructions of the physical environment into the VR application.
Tao and Lopes \cite{tao2022integrating} show that potential real-world distractions can be integrated in VR to improve presence.
Therefore, Simeone et al. \cite{simeone2016vr} used a depth camera to detect and display bystanders' positions. Displaying bystanders is especially helpful, as  O'Hagen et al.~\cite{o2020reality} found that users can feel uncomfortable when knowing that bystanders are present without being aware of their position.
Other research focuses on introducing smartphones into the virtual environment by showing a video pass-through of the smartphone and the users hands~\cite{alaee2018user}, using a camera stream and screenshots sent from the phone to the VR device~\cite{desai2017window}, or replacing the phone and hands by a virtual representation in VR~\cite{bai2021bringing}.

In the video recordings that we obtained from Biener et al.~\cite{biener2022quantifying}, the participants were using an off-the-shelf \deletemarker{Oculus Quest 2}\newmarker{HMD} without any of the previously mentioned advanced techniques.
\deletemarker{The setup provided the possibility to exchange the virtual environment with a video pass-through of the physical world and it could also display a virtual representation of the desk and keyboard and a video pass-through of the hands while typing.}
\newmarker{Besides seeing a virtual representation of the keyboard and a video pass-through of their hands, they could enable video pass-through to see the physical world.}
Therefore, it gives us the opportunity to observe how participants behave in a baseline VR environment and how they handled situations in which users had to interact with bystanders or physical objects without the above-mentioned advanced techniques.

\subsection{Prolonged Studies in AR and VR}
\label{sec:prolongedStudies}
To better understand and observe all effects of using VR, it is advisable to conduct studies with a longer duration.
There are several studies that looked at the prolonged use of VR or AR devices.
For example, Steinicke and Bruder~\cite{steinicke2014self} conducted a self experiment where one person spend 24 hours working, eating, sleeping and entertaining himself in VR. 
Nordahl et al.~\cite{nordahl201912} reported on two participants who used VR for 12 hours.
\deletemarker{Using AR, }Lu et al.~\cite{lu2021evaluating} reported on an in-the-wild study in which participants used their glanceable AR prototype for three days 
\deletemarker{. The study revealed}\newmarker{concluding} that the main issue that would keep the participants from using this system daily is the form factor of the glasses. Grubert et al.~\cite{grubert2010extended} observed user of AR HMDs in an order processing task for four hours. While the work efficiency of users increased, they also reported higher visual discomfort.

Guo et al.~\cite{guo2019mixed, guo2019evaluation, guo2020exploring} and Shen et al.~\cite{shen2019mental} report on a study where 27 participants worked in a virtual and a physical office environment for eight hours each. They focused on emotional and physiological needs~\cite{guo2019mixed} during short and long-term use of VR, but also on visual fatigue and physical discomfort~\cite{guo2020exploring}. They found that the weight and form factor of the HMD resulted in higher physical discomfort in VR compared to the physical condition. Yet, there are no further insights into how participants coped with wearing the HMD.

So far, the longest \deletemarker{study about wearing virtual reality HMDs} \newmarker{VR study} was reported by Biener et al.~\cite{biener2022quantifying} \newmarker{who compared working in VR to working in a regular physical environment for five days each}.
\deletemarker{They conducted a study where participants worked in a replicated VR environment for five working days and compared it to working for five working days in a regular physical environment.} 
They reported a wide range of measures \deletemarker{such as} \newmarker{(}task load, usability, flow, productivity, frustration, positive/negative affect, wellbeing, anxiety, simulator sickness, visual fatigue, heart rate, break times, \deletemarker{and} typing speed\newmarker{)} \deletemarker{which were} taken at regular time intervals\deletemarker{ during the study}. Yet, this paper did not closely examine the video data recorded during the study, which includes valuable and interesting insights in how participants use the HMD, how they interact with it and what problems they encounter. We have been granted access to this data and analyzed around 40\% of the video data to study participant behavior in more detail. This can also contribute towards a better understanding of the ergonomics of current VR devices, as these aspects should get more attention and consideration\cite{chen2021human, ciccone2023next}.


\subsection{Video Analysis Studies}
\label{sec:videoAnalysisStudies}
\deletemarker{Prior work in the context of extended reality have also analyzed videos recorded during studies to get a better understanding of participants behavior. For example, }
\newmarker{Analyzing videos recorded during studies have been used before to get a better understanding of participant behavior.}
Southgate~\cite{Southgate2020Using} analyzed types of learning behavior in virtual environments by analyzing screen capture videos, concluding that this can be a good process to understand learning behavior within an immersive VR environment.
Kang et al.~\cite{Kang2020ARMath} conducted a study where children used a mobile AR system to learn about mathematical concepts, and analyzed the videos recorded during the study to see how children used the system. Segura et al.~\cite{Segura2021Exploring} used video analysis to describe the behavior of participants while playing games in an immersive exergame platform. For handheld AR, Grubert et al.~used video analysis for a series of studies investigating the behavior of players and bystanders in public gaming~\cite{grubert2012playing, grubert2013playing} and tourism~\cite{grubert2015utility} scenarios.

In the workplace context, Hindmarsh and Heath~\cite{hindmarsh2007video} provide an overview of several video-based studies that were conducted in different workplaces such as call centers, mobile offices or control centers. They argue that videos provide access to details of social interactions that might not be available otherwise.

\section{Methodology}
This work aims at closely analyzing the behavior of 16 participants during an experiment (6 female, 10 male, $m=29.31$ years, $sd=5.52$, ranging from 22--38 years). Throughout this paper we abbreviate arithmetic mean as $m$ and standard deviation as $sd$. All participants were university employees or researchers. Two participants had no previous experience with VR, six only slight, two moderate, four substantial, and two extensive experience. 
The participants were recorded wearing \deletemarker{HMDs}\newmarker{an off-the-shelf Oculus Quest 2} while working for five consecutive days (\textsc{vr} condition) and also for another five without an HMD (\textsc{physical} condition), as specified in Biener et al.~\cite{biener2022quantifying}. 
\newmarker{The ethics committee of Coburg University approved this study, which took place in quiet lab areas.}
\newmarker{The VR setup provided the possibility to exchange the virtual environment with a video pass-through of the physical world and participants could see a virtual representation of the physical keyboard and a video pass-through of their hands. Otherwise both conditions were kept as similar as possible.}
Each day, participants worked for eight hours, with a mandatory 45-minute lunch break after four hours. 
\deletemarker{Participants were not asked to perform any predefined study tasks, instead they}\newmarker{Instead of a predefined study task, participants} carried out their own everyday work tasks. 
The videos we use in this study were recorded during this previously mentioned study~\cite{biener2022quantifying} using a webcam \newmarker{facing the participants and recording their face and parts of their upper body}.
Similar to prior research involving video analysis \cite{Southgate2020Using, Kang2020ARMath} we used open and axial coding.
Six people were involved in watching and annotating interesting behavior in the videos (Annotator A, B, C, D, E, and F). 
All annotators had a computer science background, either as students or as employees in our lab.
As a first step, annotator A and B were skimming through one \textsc{vr} video for each participant, taking notes on possible codes for relevant behavior.
Thereafter they discussed and concluded on a first joint codebook.
Both annotator A and B used this codebook to code four hours of one video during which there were multiple iterations of discussion to extend and refine the codebook. 

The codebook for the \textsc{physical} videos was derived directly from the \textsc{vr} codebook, 
\deletemarker{as the goal for annotating the \textsc{physical} videos was}
\newmarker{to allow us }
to compare participants' behavior in \textsc{vr} with their behavior in the \textsc{physical} condition. Therefore, annotator A and B discussed which codes from \textsc{vr} would also be applicable for \textsc{physical}, as well as which additional codes were needed for \textsc{physical}. 
\deletemarker{For codes involving the HMD, we substituted them} 
\newmarker{Codes involving the HMD were substituted }
with regular glasses, if participants were wearing any. However, actions that were inapplicable to normal glasses were removed, such as managing cables or using controllers. 
\deletemarker{All codes were grouped into categories which are explained in section \ref{sec:results}.}
\newmarker{All codes were grouped into categories as described in section \ref{sec:results}.}

The study by Biener et al.~\cite{biener2022quantifying} provided us with around 698 hours of video material for the \textsc{vr} condition and around 702 hours for the \textsc{physical} condition.
The time demand for labeling one video was very high as annotators required about one hour for processing one hour of video material.
Therefore, we decided to annotate all \textsc{vr} videos for day 1, 3, and 5 and the \textsc{physical} videos only for day 1. As participants were already familiar with using a standard desktop setup for work, we were not expecting a change of behavior over time for the \textsc{physical} videos. 
This means we annotated 48 \textsc{vr} videos (420 hours)  and 16 \textsc{physical} videos (139 hours) for a total of 559 hours. In the end, annotator A and B finished eight, annotator C and D 16, annotator E 12, and annotator F four videos. The amount of completed videos varied between annotators, depending on their available time and\deletemarker{ their speed which was strongly influenced by} the number of events in each video.

\subsection{Inter-Coder Reliability}
\label{sec:Inter-CoderReliability}
To ensure a high consistency among annotations, annotator A and B first explained the codebook to the other annotators and thereafter each annotator annotated a 30-minute training video, \deletemarker{which was created through a cut of}\newmarker{extracted from} one of the \textsc{vr} videos. The results from each annotator were compared to the results from annotator A, by computing the $F$-score~\cite{hripcsak2005agreement}: 
 \begin{equation}
     F=\frac{(1+\beta^2) \cdot recall \cdot precision}{(\beta^2 \cdot precision)+recall}. 
 \end{equation}

As there was no indication that we should weight recall over precision or vice versa, the weight $\beta$ was set to $1$.
Recall is the \newmarker{number of} overlap between two annotators divided by all annotations of the first annotator\deletemarker{ with regards to one code}, 
and precision is the \newmarker{number of} overlap between two annotators divided by all annotations of the second annotator\deletemarker{ with regards to one code}.

The $F$-score was first calculated for each code and then averaged among all codes to arrive at a single value comparing the two annotators. The average $F$-score for the training video was 0.77 (sd=0.09).
The annotations for the training video were also checked manually to identify any problems, which were then discussed among annotators, as some actions could not be unambiguously assigned to one \deletemarker{of the codes}\newmarker{code}. 

After discussing the training video, each annotator was assigned videos of several participants. 
Upon starting annotating videos of a new participant, a random 30-minute-video-section of that participant was annotated by two different annotators, one of which was always annotator A, and then checked for inter-coder reliability. 
The average $F$-score for these tests was 0.81 (sd=0.09).
After discussing potential issues, the corresponding annotator proceeded to annotate all videos of this participant, including short discussions with other annotators about unclear events. Therefore, all four videos (three \textsc{vr} and one \textsc{physical}) \deletemarker{for every single participant}\newmarker{of one participant} were annotated by a single annotator to allow for the highest practically possible consistency in the annotation process.

\subsection{Labeling Process}
We used ELAN~\cite{ELANsoftware,brugman2004annotating} to code the videos.
We set it up to have multiple tiers to add codes, as it was possible that labels would overlap. For example, participants could be talking while rubbing their eyes. 
Annotators usually watched the video at twice the original speed and stopped for adding annotations by dragging the mouse in the timeline to mark the corresponding time-span and then selecting the appropriate annotation from a predefined dictionary with all the codes.
HMD-related events were generally annotated from the moment the participant was touching the HMD to the moment when the hands stopped touching it. 
\deletemarker{The act of }Taking off the headset was annotated from the moment the participant touched the HMD until the HMD no longer touched the head of the participant and the other way round for putting the HMD on. 
For other events, the annotation was added from the point where it was apparent to the annotator that the event started until the point when the event concluded. This included observing arm movements to estimate the start and end of actions, such as using the controller \deletemarker{when the controller is not visible for the entire duration of an event}\newmarker{while it is not visible in the video}.


\subsection{Statistical Analysis}
\label{sec:results}
We grouped all codes into 17 categories. 
We analyzed the occurrence of each category in \textsc{vr} using a repeated measures ANOVA. As in prior work~\cite{biener2022quantifying} we analyzed \textsc{day} (\textsc{day 1}, \textsc{day 3}, \textsc{day 5}) and \textsc{time} (\textsc{morning}, \textsc{afternoon}) as independent variables.
As the average number and the total duration of the events per hour are the dependent variables, we divided the total number and duration during each time period (\textsc{morning}, \textsc{afternoon}) by the length of this time period, which was usually around four hours and 22 minutes (four hours of work plus half of the 45-minute break).
To ensure the robustness of the ANOVA, even with data that is not normally distributed~\cite{blanca2023non}, we used Greenhouse-Geisser correction whenever the sphericity assumption was violated.
We applied Bonferroni-correction to all post-hoc tests involving multiple comparisons.

For categories that are also sensible in \textsc{physical}, we compared \textsc{day 5} of \textsc{vr}, in which participants were already more familiar with the HMD, to \textsc{day 1} of \textsc{physical} using a repeated measures ANOVA with the independent variables \textsc{interface} (\textsc{vr}, \textsc{physical}) and \textsc{time} (\textsc{morning}, \textsc{afternoon}). 

\deletemarker{The analysis was run on the complete data set. }In addition, we also ran a separate repeated measures ANOVAs for each measure with gender as a between-subjects factor to test for gender differences. We did not run separate analysis for each gender, as the two groups are very small. However, the descriptive data and trends of both individual groups (male, female) is in line with the significant differences reported in the results.

\begin{figure*}[t]
	\centering 
	\includegraphics[width=1.0\linewidth]{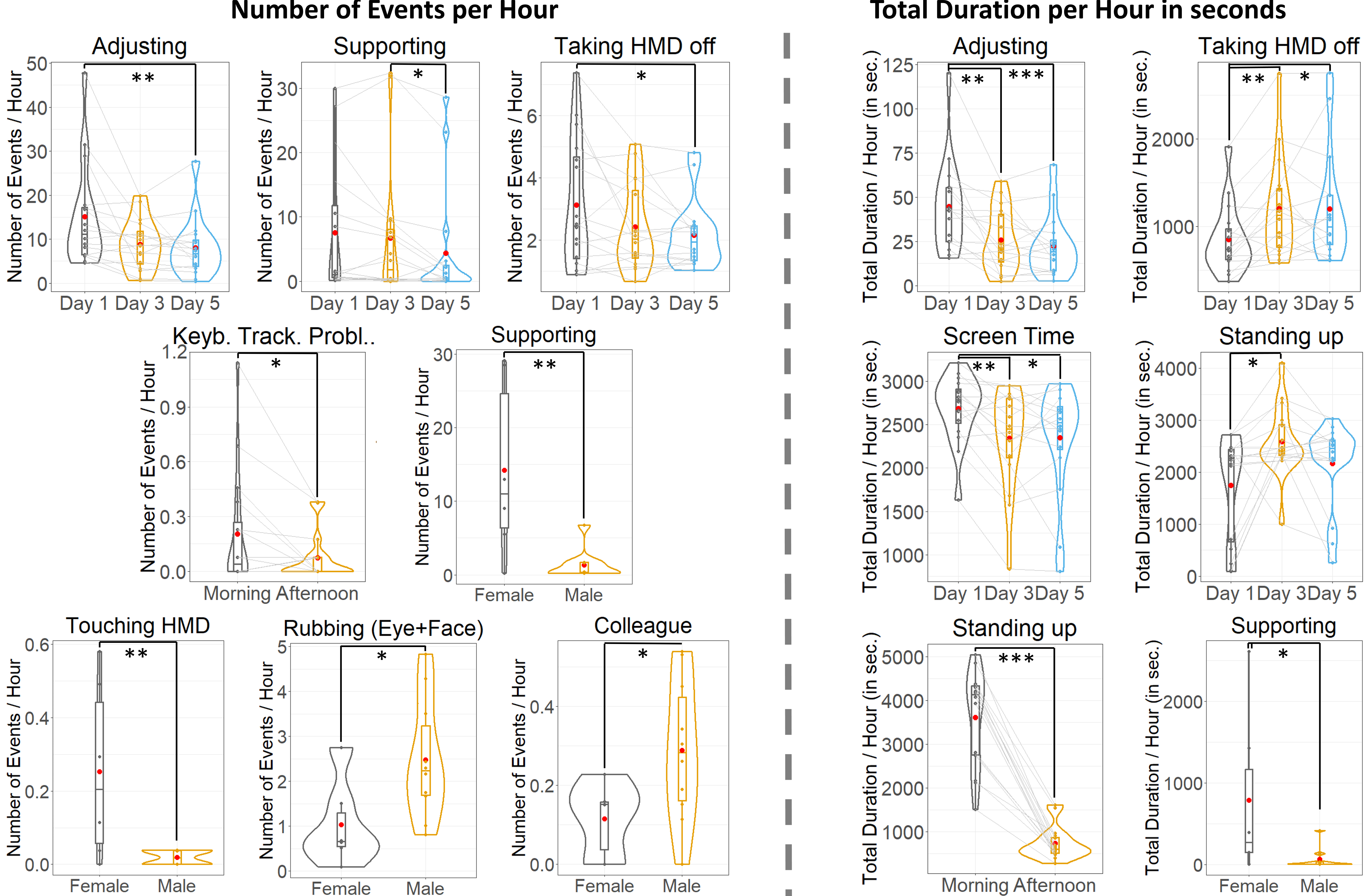}	
 \caption{Violin plots, showing the number and total duration of events per hour in \textsc{vr} separated by \textsc{day}, \textsc{time} and gender. The red dot displays the arithmetic mean. Stars indicate the level of significance in post-hoc tests: $*<0.05$, $**<0.01$, $***<0.001$.}
	\label{fig:VRplots}
\end{figure*}

\section{Results}
The results are clustered into 17 categories, including `Other'. All significant ANOVA results can be found in table 1 and 2.

\paragraph{\textbf{Adjusting}}
This category describes all events where the participants adjusted their HMD, by moving it or making any part tighter or looser.
This includes: adjusting the HMD from below by pushing it upwards; adjusting from the side with one hand; adjusting with both hands; adjusting with one hand or two hands at the back of the head; adjusting, moving or touching the fitting strap on top of the head; making it tighter or looser by rotating the fitting wheel at the back of the head; and fixing the face pad by moving the finger between the face and the headset. 
A representative visualization of such events is depicted in Fig. \ref{fig:teaser} (a).

Analyzing how often the participants adjusted the HMD per hour showed that this number was significantly influenced by the \textsc{day}\deletemarker{($F(1.42,21.3)=8.06$, $p=0.005$, $\eta^2_p=0.35$)}.
Post-hoc tests showed that there were significantly less adjusting actions on \textsc{day 5} ($m=8.06$, $sd=7.01$) compared to \textsc{day 1} ($m=15.07$, $sd=12.64$, $p=0.010$).
Similarly, the average time per hour spent on adjusting the HMD was significantly influenced by the \textsc{day}\deletemarker{($F(2,30)=14.93$, $p<0.001$, $\eta^2_p=0.5$)}  
and post-hoc tests showed it was significantly higher on \textsc{day 1} ($m=44.88~s$, $sd=29.24$) compared to \textsc{day 3} ($m=25.7~s$, $sd=21.91$, $p=0.008$) and \textsc{day 5} ($m=22.3~s$, $sd=17.87$, $p<0.001$).
The number and duration of adjusting events can be seen in Fig. \ref{fig:VRplots}.
We did not find significant differences between genders for this measure.
Examination of the frequency of the individual codes used to label the videos revealed that 34.4\% of them was using two hands, which was the most common, followed by one hand from the side at 25.3\%.

There were five participants who wore glasses during the study. Comparing their adjusting behavior in \textsc{vr} with \textsc{physical} revealed that two of them adjusted their regular glasses more often than the HMD (one participant around 80\% more; another more than ten times as much), while the other participants did it less (96\%, 67\% and 30\% less).
For all except one participant, the time spent on adjusting per hour on average was less in \textsc{physical}.

\begin{table}[t]
    \centering 
    \small
    \begingroup
    \setlength{\tabcolsep}{5pt}

    \caption{Significant RM-ANVOA results describing changes within \textsc{vr}.}

    \begin{tabular}{|>{\centering}m{1.5cm}|>{\centering}m{0.7cm}|>{\centering}m{0.7cm}|>{\centering}m{0.7cm}|>{\centering}m{0.9cm}|c|}
    
        \multicolumn{6}{c}{\small\bfseries\textbf{Number of Events per Hour}} \\

        \hline 
        Ind. Variable & $df_{1}$ & $df_{2}$ & F & p &  $\eta^2_p$   \\ 
        \hline 

        \multicolumn{6}{|c|}{Adjusting}\\
        \hline
        \textsc{day} &  $1.42$ & $21.3$  & $8.06$  & $0.005$  & $0.35$       \\
        \hline 

        \multicolumn{6}{|c|}{Supporting}\\
        \hline
        \textsc{day} &  $2$ & $30$  & $4.8$  & $0.016$  & $0.24$       \\
        \hline 
        \textsc{gender} &  $1$ & $14$  & $11.4$  & $0.008$  & $0.45$       \\
        \hline

        \multicolumn{6}{|c|}{Taking HMD Off}\\
        \hline
        \textsc{day} &  $2$ & $30$  & $5.05$  & $0.013$  & $0.25$       \\
        \hline 

        \multicolumn{6}{|c|}{Touching HMD}\\
        \hline
        \textsc{gender} &  $1$ & $14$  & $9.63$  & $0.008$  & $0.41$       \\
        \hline 

        \multicolumn{6}{|c|}{Keyboard Tracking Problem}\\
        \hline
        \textsc{time} &  $1$ & $15$  & $5.08$  & $0.04$  & $0.25$       \\
        \hline 

        \multicolumn{6}{|c|}{Rubbing Eyes or Face}\\
        \hline
        \textsc{gender} &  $1$ & $14$  & $5.28$  & $0.057$  & $0.27$       \\
        \hline 

        \multicolumn{6}{|c|}{Colleague}\\
        \hline
        \textsc{gender} &  $1$ & $14$  & $4.69$  & $0.048$  & $0.25$       \\
        \hline 
        
    \end{tabular}

    \begin{tabular}{|>{\centering}m{1.5cm}|>{\centering}m{0.7cm}|>{\centering}m{0.7cm}|>{\centering}m{0.7cm}|>{\centering}m{0.9cm}|c|}
    
        \multicolumn{6}{c}{\small\bfseries\textbf{Total Duration per Hour}} \\

        \hline 
        Ind. Variable & d$f_{1}$ & d$f_{2}$ & F & p &  $\eta^2_p$   \\ 
        \hline 

        \multicolumn{6}{|c|}{Adjusting}\\
        \hline
        \textsc{day} &  $2$ & $30$  & $14.93$  & $<0.001$  & $0.5$       \\
        \hline 

        \multicolumn{6}{|c|}{Supporting}\\
        \hline
        \textsc{gender} &  $1$ & $14$  & $5.06$  & $0.041$  & $0.27$       \\
        \hline 

        \multicolumn{6}{|c|}{Taking HMD Off}\\
        \hline
        \textsc{day} &  $2$ & $30$  & $8.87$  & $<0.001$  & $0.37$       \\
        \hline 

        \multicolumn{6}{|c|}{Screen Time}\\
        \hline
        \textsc{day} &  $2$ & $30$  & $8.31$  & $0.001$  & $0.36$       \\
        \hline 

        \multicolumn{6}{|c|}{Standing up}\\
        \hline
        \textsc{day} &  $2$ & $30$  & $6.15$  & $0.006$  & $0.29$       \\
        \hline
        \textsc{time} &  $1$ & $15$  & $136.79$  & $<0.001$  & $0.9$       \\
        \hline
     
    \end{tabular}

    \endgroup
    \label{tab:anovaVR}
\end{table}

\paragraph{\textbf{Supporting}} 
This category describes all events where participants held the HMD in a way that suggests the purpose was to hold the HMD in a more comfortable or correct position. In contrast to the adjust-category, the HMD is barely moving and supporting generally lasts longer than adjusting.
This category includes: supporting with one hand or both hands from below; and supporting it with one hand or both hands from the side. 
A representative visualization of such events is depicted in Fig. \ref{fig:teaser} (b). 

Analyzing how often the participants supported the HMD per hour showed that this number was significantly influenced by the \textsc{day}\deletemarker{($F(2,30)=4.8$, $p=0.016$, $\eta^2_p=0.24$)}.  
Post hoc-tests showed there were significantly less supporting actions on \textsc{day 5} ($m=4.38$, $sd=9.07$) compared to \textsc{day 3} ($m=6.67$, $sd=10.85$) ($p=0.041$) as displayed in Fig. \ref{fig:VRplots}.
However, we could not find a significant influence of \textsc{day} or \textsc{time} on the total duration of supporting events even though it was decreasing from \textsc{day 1} ($m=385.76~s$, $sd=695.3$) to \textsc{day 5} ($m=308.78~s$, $sd=730.70$).
An analysis of gender as a between-subjects factor revealed a significant difference in the number of supporting events\deletemarker{($F(1,14)=11.4$, $p=0.008$, $\eta^2_p=0.45$)}, such that it occurred around ten times more often for female ($m=14.24$, $sd=12.69$) than male ($m=1.34$, $sd=2.48$) participants. We also found a significant difference between the duration of supporting events\deletemarker{($F(1,14)=5.06$, $p=0.041$, $\eta^2_p=0.27$)}, which was around 12 times longer for females ($m=790.21~s$, $sd=995.95$) than males ($m=62.9~s$, $sd=174.42$). This is also displayed in Fig. \ref{fig:VRplots}.
This indicates that in general female participants were supporting the HMD a lot more than male participants. However, it is also notable, that the standard deviation of these measures for females is much higher than for males.


We also observed than in the beginning of the study, the number and the total duration of supporting events was much higher (average around 30 times higher) for six participants compared to the others. 
For all of them except P15 and P10, the number and duration of supporting events decreased slightly during the week.
When asked about the reasons for supporting the HMD, P04, P08, P09 and P10 mentioned that this helped them to fix the HMD in a position where they get the sharpest image. P04 mentioned this was especially difficult for her as she wore regular glasses underneath the HMD.
P08 and P15 also mentioned that without using their hands, they would need to use their cheek muscles to lift the HMD.
P15 said that making the head-strap of the HMD tighter caused her to feel dizzy and P14 and P10 did not like the pressure on their face, so they kept it rather loose and supported the HMD with their hands instead.
P09 mentioned the HMD bothered him because he is not used to wearing something on his head.
However, it was visible in the video that the participants had to stop supporting the HMD at times to be able to do things such as typing.
\deletemarker{An analysis of}\newmarker{Analyzing} the frequency of individual codes also revealed that for more than 91\% of supporting events participants used one hand only.
We could not find indications that wearing glasses or a lack of VR experience would lead to a high number of supporting events. 
However, \deletemarker{as indicated by the significant gender differences}\newmarker{matching the significant gender differences}, five out of these six participants were female, while only six out of the 16 participants in the study were female.

\paragraph{\textbf{Taking HMD Off}}
To analyze how often and for how long participants took the HMD off, we annotated parts in the video where participants take the HMD off and put it back on. 
In the process of these events adjusting the HMD was not labeled separately, but considered as being part of putting the headset on, or taking it off. 

\deletemarker{We know how often participants took off the HMD, by counting the occurrences of the take-off event, and we arrived at }The time durations for not wearing a HMD \deletemarker{by calculating}\newmarker{are} the time difference between take-off and put-on events.
Analyzing how often the participants took off the HMD per hour showed that this was significantly influenced by the \textsc{day}\deletemarker{($F(2,30)=5.05$, $p=0.013$, $\eta^2_p=0.25$)}.  
Post-hoc tests showed that participants took off the HMD significantly less on \textsc{day 5} ($m=2.17$, $sd=1.16$) compared to \textsc{day 1} ($m=3.14$, $sd=2.26$) ($p=0.04$).
For the average time per hour that participants were not wearing the HMD, we also found that this was significantly influenced by \textsc{day}\deletemarker{($F(2,30)=8.87$, $p<0.001$, $\eta^2_p=0.37$)}. 
In contrast to the number of events, post-hoc tests showed that the duration was significantly lower on \textsc{day 1} ($m=849.42~s$, $sd=492.85$) compared to \textsc{day 3} ($m=1202.78~s$, $sd=687.28$, $p=0.002$) and \textsc{day 5} ($m=1199.20~s$, $sd=643.40$, $p=0.015$). The number and duration of \deletemarker{periods in which HMDs were not worn}\newmarker{taking-off events} are displayed in Fig. \ref{fig:VRplots}.
We did not find significant differences between genders for this measure.
\deletemarker{We also checked the correlation between the average number and duration of supporting actions for each participant with the average number and duration of taking off the HMD. This did not show any significant correlations, which indicates that supporting the HMD did not significantly reduce the number or duration of take-off events.}
\newmarker{Also, we could not find any significant correlations between the average number and duration of supporting actions and the average number and duration of taking off the HMD for each participant.}

To gain a better understanding of why participants took the HMD off, we checked which other labels occurred while the HMD was not worn. The most frequent ones were talking (14.6\%), using the phone (12.4\%), standing up (12.0\%) or sitting down (11.8\%), rubbing the face (11.4\%) and drinking (10.5\%).

For the five participants who wore regular glasses in the \textsc{physical} condition, we noticed that one participant took off the regular glasses more often than the HMD while all other participants did it less and one participant never did it. The average duration per hour that they did not wear the glasses was much lower for all participants in \textsc{physical} ($m=103.84~s$, $sd=201.12$) compared to \textsc{vr} ($m=1277.58~s$, $sd=692.87$).

\begin{table}[t]
    \centering 
    \small
    \begingroup
    \setlength{\tabcolsep}{5pt}

    \caption{Significant RM-ANOVA results comparing \textsc{vr} and \textsc{physical}. }

    \begin{tabular}{|>{\centering}m{1.5cm}|>{\centering}m{0.7cm}|>{\centering}m{0.7cm}|>{\centering}m{0.7cm}|>{\centering}m{0.9cm}|c|}
    
        \multicolumn{6}{c}{\small\bfseries\textbf{Number of Events per Hour}} \\

        \hline 
        Ind. Variable & $df_{1}$ & $df_{2}$ & F & p &  $\eta^2_p$   \\ 
        \hline 

        \multicolumn{6}{|c|}{Eating or Drinking}\\
        \hline
        \textsc{interface} &  $1$ & $15$  & $7.97$  & $0.013$  & $0.45$       \\
        \hline 

        \multicolumn{6}{|c|}{Phone Call}\\
        \hline
        \textsc{time} &  $1$ & $15$  & $7.2$  & $0.02$  & $0.33$       \\
        \hline

        \multicolumn{6}{|c|}{Rubbing Eyes or Face}\\
        \hline
        \textsc{interface} &  $1$ & $15$  & $5.6$  & $0.032$  & $0.27$       \\
        \hline
        \textsc{time} &  $1$ & $15$  & $7.64$  & $0.014$  & $0.34$       \\
        \hline

        \multicolumn{6}{|c|}{Rubbing Eyes}\\
        \hline
        \textsc{interface} &  $1$ & $15$  & $11.75$  & $0.004$  & $0.44$       \\
        \hline
        \textsc{time} &  $1$ & $15$  & $9.03$  & $0.009$  & $0.38$       \\
        \hline

        \multicolumn{6}{|c|}{Physical World Activity}\\
        \hline
        \textsc{interface} &  $1$ & $15$  & $7.669$  & $0.014$  & $0.34$       \\
        \hline

    \end{tabular}

    \begin{tabular}{|>{\centering}m{1.5cm}|>{\centering}m{0.7cm}|>{\centering}m{0.7cm}|>{\centering}m{0.7cm}|>{\centering}m{0.9cm}|c|}
    
        \multicolumn{6}{c}{\small\bfseries\textbf{Total Duration per Hour}} \\

        \hline 
        Ind. Variable & d$f_{1}$ & d$f_{2}$ & F & p &  $\eta^2_p$   \\ 
        \hline 

        \multicolumn{6}{|c|}{Standing up}\\
        \hline
        \textsc{time} &  $1$ & $15$  & $66.97$  & $<0.001$  & $0.82$       \\
        \hline 

        \multicolumn{6}{|c|}{Rubbing Eye}\\
        \hline
        \textsc{time} &  $1$ & $15$  & $10.33$  & $0.006$  & $0.41$       \\
        \hline 
     
    \end{tabular}

    \endgroup
    \label{tab:anovaVRvsPHYS}
\end{table}

\paragraph{\textbf{Screen Time}}
We calculated screen time in \textsc{vr} by including all the times when the participants were wearing the HMD, not including all the times in which they were taking the HMD halfway off, using the controller or tapping on the HMD, encountering keyboard tracking problems or were reading, writing or using their phone.
We found a significant main effect of \textsc{day} on the screen time per hour in \textsc{vr}\deletemarker{($F(2,30)=8.31$, $p=0.001$, $\eta^2_p=0.36$)}, such that the screen time was higher on \textsc{day 1} ($m=2684.62~s$, $sd=494.58$) compared to \textsc{day 3} ($m=2343.18~s$, $sd=686.0$, $p=0.004$) and \textsc{day 5} ($m=2348.58~s$, $sd=643.88$, $p=0.017$). 
We did not find significant differences between genders.
The screen time for all three \textsc{vr} days is shown in Fig. \ref{fig:VRplots}.

In \textsc{physical} we included two additional codes that mark events where the participant turned away from the screen and turned back to look at the screen to compare the screen time in \textsc{vr} and \textsc{physical}. This code could not be reliably used for \textsc{vr}, because it is not clear from the videos where the participants had placed their virtual monitor and therefore if they are looking at it or not.
Screen time in \textsc{physical} was then calculated similarly to \textsc{vr}, as the time that participants were looking at the screen not including all times in which they were reading, writing or using the phone.
Comparing \textsc{vr} with \textsc{physical} showed an interaction effect of \textsc{day} and \textsc{time} but post-hoc tests did not show significant differences.

\paragraph{\textbf{Touching HMD}}
We introduced one code which labels events where participants are touching the HMD without adjusting or supporting it or doing any other purposeful action.

Overall, this is a very rare event which did not even occur for 6 participants. 
Therefore, we could also not find any significant influence of \textsc{time} or \textsc{day} on the number or duration of such events.
However, looking at gender as a between-subjects factor revealed a significant difference in number of touching events\deletemarker{($F(1,14)=9.63$, $p=0.008$, $\eta^2_p=0.41$)} such that it occurred more often for females ($m=0.25$, $sd=0.57$) than males ($m=0.02$, $sd=0.06$) as is displayed in Fig. \ref{fig:VRplots}. No difference between genders was detected for the total time spent on such events.

From what the annotators can estimate from the videos, four participants touched the HMD to check its position, five were just touching it for no apparent reason, three moved their fingers along the HMD, one was tapping on both sides, and one participant was briefly touching it with a second hand while adjusting it.

\paragraph{\textbf{Handling Cable}}
This category describes all events related to the power-cable that was usually plugged in to the HMD during the study. It can either be that the participants are purposefully touching the cable, for example to move it out of the way or that the participants take out or plug in the cable.
A representative visualization of such events is depicted in Fig. \ref{fig:teaser} (c).

This is also a rather rare event that only occurs in around 67\% of the analyzed videos.
We did not find any significant influence of \textsc{time} or \textsc{day} on the number or duration of such events nor any differences between genders.
Looking at the labeled sections, we noticed that six participants were moving the cable behind their arm, six participants checked that it is properly plugged in and six participants simply tried to move it to the side a little.
P01 seemed to struggle with the weight of the cable pulling the head down on one side, so she tried to find a more comfortable position by laying it across her head.
Four participants unplugged it during their workday: one of them to be able to turn around and redraw the boundary; one to stand up while wearing the HMD; and two participants seemed to take it out because it was annoying.
In six videos the cable was coming from the top hanging over a movable wall to avoid extra weight. Even though the sample size is very small, this could be a strategy to avoid cable events, as the average number of events per hour with the cable hanging down was $0.46$ ($sd=0.85$) and with the cable coming from the top only $0.10$ ($sd=0.18$).

\paragraph{\textbf{Taking HMD Halfway off}}
This category describes actions in which participants lift either the front of the HMD above eye-level with one or two hands, or lift the strap up at the back without completely removing the HMD. 
A representative visualization of such events is depicted in Fig. \ref{fig:teaser} (d).

We could not find any main effects of \textsc{day} or \textsc{time} on the number or duration of such actions, nor significant gender differences.
Analyzing the frequency of the individual codes revealed that 74\% of these actions were performed with one hand and 25\% with two hands in the front and 1\% in the back.
To get an idea of why participants took the HMD halfway off, we checked which other labels overlapped with it. The most common ones were drinking (38.0\%), rubbing the face (8.5\%), talking (8.0\%), rubbing the eyes (6.9\%), using the controller (6.1\%) and using the phone (5.2\%). By examination of the overlaps we cannot be sure that they caused the participants to take the HMD halfway off, but they can be indications.
The back of the HMD was mainly lifted to scratch the head.

\begin{figure}[t]
	\centering 
	\includegraphics[width=1.0\linewidth]{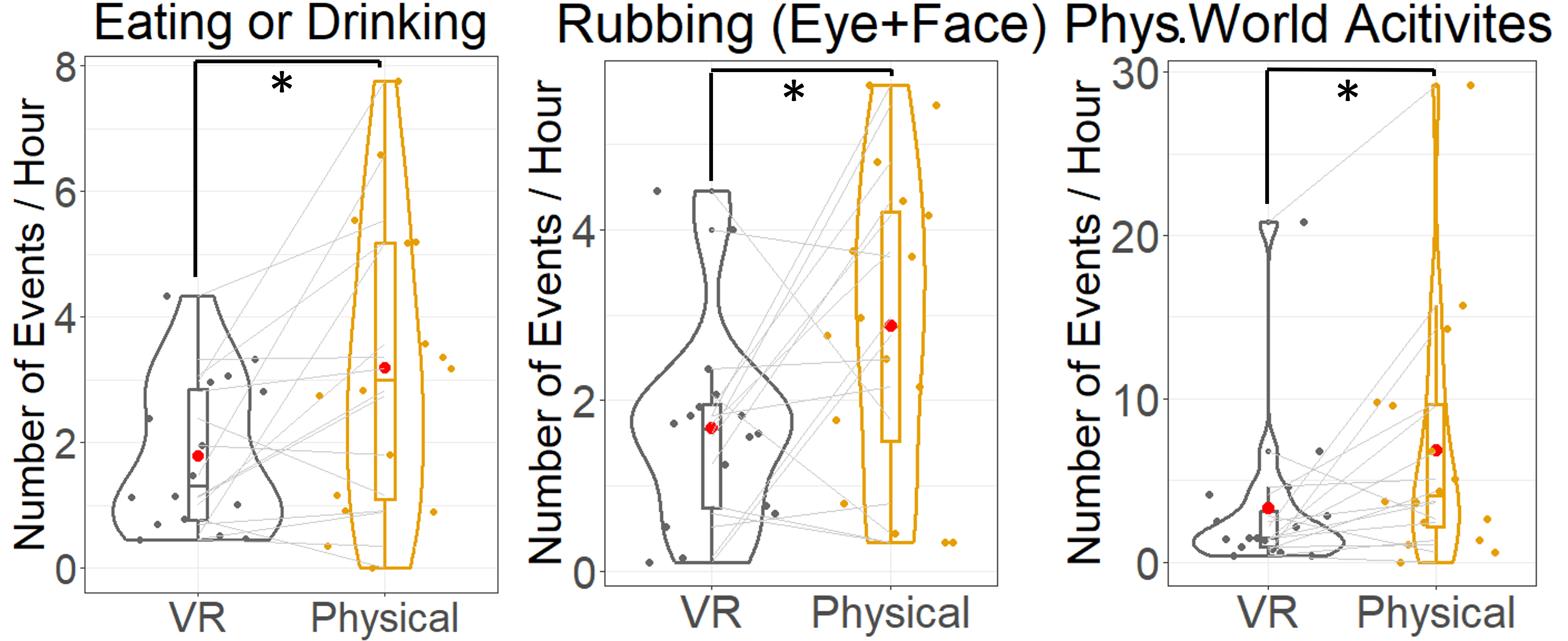}
 \caption{Violin plots showing the number of events per hour for \textsc{vr} and \textsc{physical}. The red dot displays the arithmetic mean. Stars indicate the level of significance in post-hoc tests: $*<0.05$, $**<0.01$, $***<0.001$.}
	\label{fig:compareVRPhysPlots}
\end{figure}

\paragraph{\textbf{Interacting with HMD}}
We captured how often the participants had to interact with the interface of the HMD by using the controller or hand gestures, by tapping on the HMD to enable or disable passthrough mode, by using the volume-buttons on the HMD or by redrawing the boundary of the guardian area, which always coincides with controller usage
\deletemarker{. A representative visualization of such events}\newmarker{which} is depicted in Fig. \ref{fig:teaser} (e).

We could not find any main effects of \textsc{day} or \textsc{time} on the number or duration of those actions nor did we find significant differences between genders.
This is also not expected since these actions are mainly mandatory to operate the system.
Looking at the individual codes, we saw that 83.9\% of the occurrences was about using the controller (or hand-gestures), 12.6\% about tapping for pass-through and 3.6\% about adjusting the volume.
It seemed like only four participants had to redraw the boundary during the study.
We observed that participants regularly used the controllers after putting the HMD on. This is probably due to the fact that after putting it back on the virtual desktop often moved a little, so they had to reposition it.


\paragraph{\textbf{Keyboard Tracking Problem}}
To get an indication on how often participants had problems with the keyboard tracking, we annotated parts of the video in which participants were obviously trying to get the keyboard tracked.

This is also a rather rare event, as it only occurs in around 42\% of the analyzed \textsc{vr} videos.
In these situations, participants generally lifted the keyboard and moved it around to try to track it again as visualized in Fig. \ref{fig:teaser} (f).
We found a significant main effect of \textsc{time} on the number of keyboard tracking problems\deletemarker{($F(1,15)=5.08$, $p=0.04$, $\eta^2_p=0.25$)}, such that there were significantly more in the \textsc{morning} ($m=0.2$, $sd=0.52$) compared to the \textsc{afternoon} ($m=0.07$, $sd=0.19$) 
\deletemarker{The number of tracking problems for all three days in \textsc{vr} is}\newmarker{as} visualized in Fig. \ref{fig:VRplots}.
We did not find significant differences between genders.

\paragraph{\textbf{Standing up}}
To analyze how often and for how long participants stood up from their chair, we annotated parts in the videos in which they \deletemarker{stood}\newmarker{got} up and sat down.

We could not detect a significant influence of \textsc{day} or \textsc{time} on the number of such events.
However, we found a significant main effect of both \textsc{day}\deletemarker{($F(2,30)=6.15$, $p=0.006$, $\eta^2_p=0.29$)} and \textsc{time}\deletemarker{($F(1,15)=136.79$, $p<0.001$, $\eta^2_p=0.9$)} on the duration\deletemarker{(time between SD and SU)} of standing per hour. 
Post-hoc tests showed that participants were standing longer on \textsc{day 3} ($m=2590.74~s$, $sd=1862.65$) compared to \textsc{day 1} ($m=1752.45~s$, $sd=1855.36$, $p=0.027$).
In addition, participants were standing significantly less during the \textsc{afternoon} ($m=729.83~s$, $sd=522.68$) compared to the \textsc{morning} ($m=3612.36~s$, $sd=1558.06$). These results are visualized in Fig. \ref{fig:VRplots}.

Comparing \textsc{vr} to \textsc{physical} revealed a significant influence of \textsc{time} \deletemarker{($F(1,15)=66.97$, $p<0.001$, $\eta^2_p=0.82$)} on the duration of standing per hour, yet no interaction effect between \textsc{interface} and \textsc{time}. 
This indicates that both in \textsc{vr} and \textsc{physical} participants stood less during the \textsc{afternoon} ($m=742~s$, $sd=513.7$) than during the \textsc{morning} ($m=3425~s$, $sd=1498.5$) (see Fig. \ref{fig:compareVRPhysPlots}).

\deletemarker{Overall, we did not find significant differences between genders for standing up events.}
\newmarker{No significant gender differences were found for such events.}

\paragraph{\textbf{Eating or Drinking}}
This category summarizes the two codes for drinking and eating, labeling the parts\deletemarker{of the videos} in which participants put drinks or food into their mouth, \deletemarker{not including}\newmarker{ignoring} longer periods of chewing.
Fig. \ref{fig:teaser} (g) depicts a representative visualization of such events.

We combined the occurrences of both codes to see how often and for how long participants were engaged with food.
In \textsc{vr}, we could not find a significant influence of \textsc{day} or \textsc{time} on the number or duration of such events. 

When comparing \textsc{physical} to \textsc{vr}, we found a significant influence of \textsc{interface} on the number of such events\deletemarker{($F(1,15)=7.97$, $p=0.013$, $\eta^2_p=0.45$)}, such that in \textsc{physical} ($m=3.19$, $sd=2.67$) participants drank and ate more frequently than in \textsc{vr} ($m=1.79$, $sd=1.31$) (Fig. \ref{fig:compareVRPhysPlots}). 
However, we did not find a significant influence on the total time spent eating or drinking per hour.\deletemarker{ which suggests that participants took longer to eat or drink in VR. This could be caused by them being more cautious.}
We did not find any significant differences in gender.
Looking at the frequency of drinking and eating individually, we found that participants were mostly drinking, accounting for 73.7\% of the occurrences in \textsc{vr} and 76.0\% \deletemarker{of the occurrences} in \textsc{physical}. 

Generally, participants used the same containers for drinking in \textsc{vr} and \textsc{physical} which included cups, glasses, and bottles.
In \textsc{vr}, for the majority (92.2\%) of drinking and eating events and the majority (94.1\%) of time spend eating, participants did not take off the HMD and of those in only 16\% did they lift the HMD (take halfway off), and in only 10\% of the cases did they support the HMD.
We noticed that P06 often did not support the HMD when drinking out of a bottle but supported or took it halfway off when using a cup.
Instead of supporting the HMD with the hands, P04 sometimes used the cup itself to push the HMD slightly upwards.
P02, however, at one point attempts to drink out of a glass but gives up because the HMD blocks access to the mouth.

\paragraph{\textbf{Talking}}
To detect how often, and for how long, participants were talking to others in the same room, visualized in Fig. \ref{fig:teaser} (h), we labelled such parts in the videos. As the videos were recorded without audio, we included all events in which the participants move their lips in a way that suggests they are having a conversation, not including phone calls or video calls, if this is apparent in the video.

We could not find a significant influence of \textsc{day} or \textsc{time} on the number or duration of talking events per hour, and no significant difference between \textsc{vr} and \textsc{physical} or between genders.
This is unsurprising, as talking is probably dependent on many external factors such as availability of another person.
However, we found that 78.2\% of all talking occurrences and 79.5\% of the total talking duration was done while wearing the HMD and only in 1.9\% of these did participants lift the front of the HMD (take halfway off).
\deletemarker{This indicates that it might not have been a problem for the participants to talk to people while not being able to see them.}

\paragraph{\textbf{Phone Call}}
We introduced one code for annotating all parts of the video in which the participant is talking on the phone or in a video call such as visualized in Fig. \ref{fig:teaser} i).

This is a rare event, as it only occurs in around 40\% of the analyzed \textsc{vr} videos.
We did not find any significant influence of \textsc{day} or \textsc{time} on the number or duration of phone calls and also no significant influence of \textsc{interface} or gender. However, when comparing \textsc{vr} and \textsc{physical} there was a significant main effect of \textsc{time}\deletemarker{($F(1,15)=7.2$, $p=0.02$, $\eta^2_p=0.33$)}, such that there were more phone calls in the \textsc{morning} ($m=0.25$, $sd=0.42$) compared to the \textsc{afternoon} ($m=0.15$, $sd=0.33$).
However, we have to be cautious here, as the frequency of phone calls are probably much more influenced by other people than the participants experience during the \textsc{vr} condition.

We could observe that 90.9\% of all phone call occurrences and 94.4\% of the total phone call duration occurred while wearing the HMD and in only 7\% of these instances did the participant lift the front of the HMD (take it halfway off).
P04 once removed the HMD after answering the call. Yet, at another time she put the HMD back on after starting a call. P11 removed the HMD to answer the call and then afterwards immediately put it back on. From the videos it appears participants were both answering and starting calls.

\paragraph{\textbf{Rubbing Eyes or Face}}
This category combines instances in which participants are rubbing their eyes or face whenever these actions are very obvious, not including events in which the participants very shortly, and probably involuntarily, touch their faces.
A representative visualization of such events is depicted in Fig. \ref{fig:teaser} (j).

We did not observe any significant influence of \textsc{day} or \textsc{time} on the number or duration of such actions in \textsc{vr}.
An analysis of gender as a between-subjects factor revealed a significant difference in the number of instances of rubbing eyes or faces\deletemarker{($F(1,14)=5.28$, $p=0.037$, $\eta^2_p=0.27$)}, which was about twice as high for males ($m=2.48$, $sd=1.82$) than for females ($m=1.03$, $sd=1.14$), yet no difference was detected regarding the total duration. 
Significant differences are visualized in Fig. \ref{fig:VRplots}.

Also, when comparing \textsc{vr} and \textsc{physical}, we found a significant main effect of \textsc{interface}\deletemarker{($F(1,15)=5.6$, $p=0.032$, $\eta^2_p=0.27$)} and \textsc{time}\deletemarker{($F(1,15)=7.64$, $p=0.014$, $\eta^2_p=0.34$)}
on the number of such actions 
, but no effect on the duration. 
These effects indicate that participants rubbed their eyes and faces more in \textsc{physical} ($m=2.87$, $sd=2.25$) than \textsc{vr} ($m=1.68$, $sd=1.44$) (see Fig. \ref{fig:compareVRPhysPlots}) and that they did this more often in the \textsc{morning} ($m=2.75$, $sd=2.36$) compared to the \textsc{afternoon} ($m=1.8$, $sd=1.36$). 

An analysis of the frequency of individual codes shows that in \textsc{vr} 67.1\% were about rubbing the face and 32.9\% about rubbing eyes, while, in \textsc{physical} rubbing the eyes had a portion of 59.0\%.

Therefore, we compared \textsc{vr} and \textsc{physical} also for both individual codes.
There were significantly more eye rubbing events in \textsc{physical} ($m=1.7$, $sd=1.68$) compared to \textsc{vr} ($m=0.58$, $sd=0.613$)\deletemarker{($F(1,15)=11.75$, $p=0.004$, $\eta^2_p=0.44$)},
but no significant difference for the duration\deletemarker{ per hour between \textsc{vr} and \textsc{physical}} or between genders. 
Yet, there were significantly more eye rubbing events in the \textsc{morning} ($m=1.45$, $sd=1.66$) compared to the \textsc{afternoon} ($m=0.83$, $sd=0.94$)\deletemarker{($F(1,15)=9.03$, $p=0.009$, $\eta^2_p=0.38$)} and significantly more time spent on eye rubbing in the \textsc{morning} ($m=4.82~s$, $sd=5.68$) compared to the \textsc{afternoon} ($m=2.74~s$, $sd=3.06$)\deletemarker{($F(1,15)=10.33$, $p=0.006$, $\eta^2_p=0.41$)}.
No significant differences were found regarding rubbing the face.

\paragraph{\textbf{Physical World Activities}}
\deletemarker{We were also interested in how often and for how long participants concern themselves with things outside of the virtual world. Therefore,}
\newmarker{To explore how often and for how long participants concern themselves with things outside of the virtual world,} this category combines codes that describe events where the annotators believe the participants were reading something outside of VR, writing something outside of VR, using a smartphone \newmarker{(Fig. \ref{fig:teaser} (k))} or otherwise peeking under the HMD to see something.
\deletemarker{A representative visualization of such events is depicted in Fig. \ref{fig:teaser} (k).}
These events can occur together with previously mentioned events such as taking the HMD (halfway) off.

We found no significant influence of \textsc{day} or \textsc{time} on the number or duration of events in this category.
Comparing \textsc{vr} and \textsc{physical} indicated a significant main effect of \textsc{interface}\deletemarker{($F(1,15)=7.669$, $p=0.014$, $\eta^2_p=0.34$)} such that physical world activities where significantly more frequent in \textsc{physical} ($m=6.88$, $sd=8.40$) than in \textsc{vr} ($m=3.28$, $sd=4.97$) (see Fig. \ref{fig:compareVRPhysPlots}). However, no significant differences have been found regarding the total time spent on such actions.\deletemarker{, which could indicate that in \textsc{physical} they were more accessible for spontaneous actions.}
We also did not find significant differences between gender.

An analysis of the frequency of individual events in \textsc{vr} showed that in 64.0\% of the cases participant were using their phone.
In 20.1\% of the cases they were probably reading and in 10.0\% of the cases they were writing something.
However, in \textsc{physical} only 46.6\% of the actions were related to the phone and 35.6\% to writing and 17.7\% to reading.
We also found that 92.3\% of phone usage in \textsc{vr} happened while wearing the HMD and only in 7\% of these cases did participants lift the HMD in the front (taking it halfway off).


\paragraph{\textbf{Colleague}}
To detect if an event was likely triggered by a colleague or another person, we used this code in combination with any other code to label such events. For example, the participant could be taking off the HMD when a colleague approaches.

We could not find any statistically significant influence of \textsc{day} or \textsc{time}. 
An analysis of gender as a between-subjects factor revealed a significant difference in number of events triggered by colleagues\deletemarker{($F(1,14)=4.69$, $p=0.048$, $\eta^2_p=0.25$)} which was higher for males ($m=0.29$, $sd=0.34$) than females ($m=0.11$, $sd=0.19$) as depicted in Fig. \ref{fig:VRplots}. However, this measure heavily depends on the coworkers, their availability and relationship with the participant.

Checking which actions were triggered by colleagues in \textsc{vr}, showed that 37.4\% were taking off the HMD, 26.6\% were talking, 8.6\% were taking the HMD halfway off, and 5.7\% were taking the headphones off.
In \textsc{physical}, most common actions were to turn away from (81\%) or towards (6\%) the screen, taking the headphones off (5\%) and talking (3\%).
These values can give some indications on how users react when approached by a colleague.

\paragraph{\textbf{Other}}
We also added a code\deletemarker{ (UNKNOWN)} to label any behavior that the annotators might find interesting that do not match any of the previously mentioned codes.
For example, sometimes the washable face-pad that we used in the study got out of place, so we saw five participants adjusting the face pad while not wearing the HMD. 
Additionally, we could observe three participants cleaning the HMD during the study.

\section{Discussion}
The results provide insights into the behavioral patterns of the participants as they acclimated to the experience of working in VR.
Exposure to unfamiliar technology is well known to demand an initial period of familiarization and adaptation, as also seen in studies on user experience in VR and AR \cite{kim2020systematic, kim2018revisiting}.
In this work, however, we investigated behavioral changes over a much longer timescale than typical VR user studies so far, and our observations reveal rich patterns of user behavior not previously documented.

\deletemarker{We found that the number and duration of adjusting events reduced over the five days. This}\newmarker{The observed decline of adjusting events} suggests that users need less time to fix the HMD in a comfortable position as they become more familiar with the device which highlights the need for improved tutorials for initial fitting. 
In addition, the number of supporting events \newmarker{but not the total time spent supporting} declined significantly over the week.\deletemarker{but the reduction in the total time spent supporting was not significant.} 
This could indicate that users put up with the need to support instead of repeatedly trying to wear the HMD without support. The main reasons for support, as mentioned by the participants, were to obtain a clear image and to avoid pressure on the head by having the head-strap too tight. This emphasizes the importance of ongoing efforts to improve the form factor of HMDs. 
In 95\% of the cases, the participants only supported the HMD with one hand, which made it possible to use the keyboard or mousepad with the other hand. Still, from an ergonomic point of view, this is clearly not desirable, especially for prolonged use of VR or AR HMDs.

Despite the limited number of participants, we observed some significant differences between genders, which were especially prominent in the number and duration of supporting events. Similarly, in their eight-hour-long study, Guo et al.~\cite{guo2020exploring} also noticed significant gender differences regarding visual fatigue, which was much higher for females. Yet in both cases, the number of participants was rather small, and other factors such as experience with VR devices or games, as mentioned by Guo et al.~\cite{guo2020exploring}, cannot be ruled out.
Therefore, gender differences should be investigated more closely in future research and these findings emphasize the importance of having a diverse group of participants.

\deletemarker{Similar to the adjusting and supporting events, }The frequency of taking the HMD off decreased over the five days. However, the break durations and time spent standing increased and consequently, screen time decreased. Yet, we did not detect any significant difference in screen time between \textsc{vr} and \textsc{physical}.
Also,\deletemarker{no significant correlations were found between the number or duration of supporting events and taking the HMD off, suggesting} \newmarker{our results suggest} that supporting the HMD does neither significantly increase, nor decrease the need for taking the HMD off.

\deletemarker{No significant changes were detected with regard to taking the HMD halfway off. }
Reviewing why participants might lift the HMD in the front, or even take it off completely, indicated that this was done to drink, rub the face and eyes, talk, or to use a phone. 
These are either basic needs or common actions in work environments that should not be restricted by the VR device.
Future HMDs could reduce the need for lifting the HMD when drinking or rubbing the face, by reducing the form factor or allowing easier access to the face and especially the mouth while wearing the HMD. \newmarker{One solution could be a raisable visor as offered by the HoloLens 2 \cite{hololens2}.}
When talking or using the phone users could benefit from a clearer pass-through mode or other techniques to include outside information, such as smartphones~\cite{alaee2018user, desai2017window, bai2021bringing} or parts of the surrounding environment~\cite{hartmann2019realitycheck, wang2022realitylens}.

Differences between \textsc{vr} and \textsc{physical} were found in the frequency of participants rubbing their eyes, which was significantly more frequent in \textsc{physical} and could be explained by the eyes being more accessible. The HMD could prevent participants from involuntarily rubbing their eyes as it requires more actions, such as lifting the HMD.
Also, in \textsc{physical}, participants interacted significantly more with physical objects, such as reading, writing, or using a phone \newmarker{which could indicate that they were more accessible for spontaneous actions.} \deletemarker{matching}\newmarker{This matches} the reports of Biener et al~\cite{biener2022quantifying}, who mentioned that participants missed the ability to write things down in VR.
Similarly, the frequency of eating and drinking was higher in \textsc{physical}, indicating that the HMD was too obstructive to support such activities. This was also visible in the videos as participants often supported or lifted the HMD to drink and it is also in line with participants' comments reported by Biener et al.~\cite{biener2022quantifying} about being afraid to spill something, as well as prior observations by McGill et al.~\cite{mcgill2015dose}.
On the other hand, no significant differences were found in the total duration of drinking or eating events, physical world activities, or in rubbing of eyes or faces. 
This indicates that participants did not neglect these needs, but instead did them more carefully and for longer periods, as it was not convenient to do them as often in \textsc{vr}.
Also, the HMD did not substantially restrict participants from performing actions outside VR. 
For example, 92\% of phone usage in the \textsc{VR} condition was done while wearing the HMD and participants only rarely lifted the HMD to do so.
Additionally, during 80\% of the time spent talking, 94\% of the time spent on phone calls, and 94\% of the time spent drinking or eating, participants wore the HMD. 
Still, further efforts to make such interactions more comfortable should not be neglected, such as including the phone, as mentioned above, or hinting at and visualizing bystanders \cite{simeone2016vr} or objects.
This finding sheds further light on how people adopt and appropriate technology in their daily routines, even though those routines might initially be hampered by that very technology.

Another problem could be caused by the power cable, which can pull the HMD down and become distracting when obstructing users' movements. When the HMD should be used for a time span that requires charging while working, one should think about how to comfortably adjust the cable, for example by hanging the cable from the ceiling or running it down at the back of the head rather than the side, which has already been done for newer HMDs such as Varjo Aero or Apple’s Vision Pro, or include enough breaks to charge the HMD in between uses.
Participants also seemed to have issues with the keyboard tracking\deletemarker{which they tried to solve by lifting it up and moving it around}. This happened more often in the \textsc{morning}\deletemarker{, yet it can only be speculated as to why.} \newmarker{and we can speculate that} it could be caused by different lighting conditions or because participants simply coped with not seeing the keyboard in the \textsc{afternoon}, albeit 
window blinds were used to control lighting conditions.
Taking into account the significantly fewer standing events that tended to occur in the afternoon it could also hint at participants being generally less active in the afternoon than earlier in the day.
Still, the observed problems\deletemarker{, again, suggest}\newmarker{underline} the need to further improve all components of VR HMDs, including robust object and environment tracking algorithms.

Overall, these varied observations and findings can inform the design of more comfortable HMDs for work applications, and serve as guidance for extended use of VR in work settings and risk assessments of such activities. Hardware solutions are likely required to address the trade-off between setting straps tightly to ensure a stable image and the associated discomfort arising from this tightness after extended use. Current generation HMDs, such as the Oculus Quest Pro, do offer different ergonomics compared to the Quest 2. Yet, on an anecdotal level only, we experienced that even with these changes prolonged use beyond 30--45 minutes can still lead to increased pressure on the forehead, and, subsequently, headache. Hence, it should be carefully studied if these new headset generations deliver a better long-term usage experience or not.    
Prior work has analyzed and designed workplace ergonomics by using VR as a tool \cite{whitman2004virtual, grajewski2013application}. However, less focus has been on the ergonomics of VR itself \cite{chen2021human,ciccone2023next}, or on possible solutions, such as Wentzel et al.~\cite{wentzel2020improving}, who amplified hand movements for more comfortable interactions in VR, or McGill et al.~\cite{mcgill2020expanding}, who reduced the need for head movement in large-display setups.
Software solutions may potentially be able to address other comfort issues, such as streamlining interactions to minimize user effort, prompting users towards more ergonomic virtual display configurations, and facilitating common tasks currently not easy to perform while wearing a HMD, such as locating items on the desk, using a phone, and eating or drinking.
Software-based interventions may also allow for promoting compliance with ergonomic guidance, for example by encouraging workers to take breaks after predefined time periods, using eye exercises to alleviate digital eye strain~\cite{hirzle2022understanding}, performing some form of exercise after set periods, or incorporating relaxation methods designed for VR \cite{thoondee2017usingvirtual, valtchanov2010restorative}. Although many of these design implications are intuitive, there has been limited evidence to support and motivate such remedial efforts, and, hence, we call for increased community efforts to work towards VR (and AR) experiences that can support prolonged use in various contexts.

Currently, research is often focused on brief usability evaluations. Yet, we argue that long-term usability issues require further attention. In addition, the issues of working in VR for a prolonged time, as described in this study, which have not been formally documented and described before, provide a reference against which future hardware and software improvements can be measured.

\section{Limitations and Future Work}
As mentioned in the study by Biener et al.~\cite{biener2022quantifying}, a limitation is that tasks are chosen by participants. 
This means we cannot fully eliminate the possibility that the type of tasks on different days influenced the behavior of the participants. 
Also, no VR-specific benefits were employed, as the goal of the previous study was to quantify baseline usability issues. Therefore, the behavior of participants could be different when using an optimized version of VR that, for example, facilitates certain tasks or specifically aims at increasing the well-being of the users. 
Also, only one type of HMD, the Quest 2, was investigated. Future work should also consider HMDs that vary in important aspects such as weight, weight-distribution and resolution.
Additionally, this study, as well as the prior study by Biener et al.~\cite{biener2022quantifying} is reporting the effects of using VR in the context of knowledge workers, which does not necessarily imply a generalizability to other application areas which should therefore be evaluated in future work.

Another limitation is that\deletemarker{the video dataset was collected through a webcam facing the participant and thus} we do not have a complete view of the participants' surroundings\deletemarker{. Therefore, we }\newmarker{so we can} only report what was within the camera's view. \deletemarker{For some participants (P11, P14)}\newmarker{For P11 and P14} the camera orientation and participant positioning were such that it was \newmarker{impossible}\deletemarker{difficult} to see some supporting actions or when they were using their phone\deletemarker{ as this was not visible}.

Our analysis also hinted at some gender differences, however, female participants were underrepresented in this study (six out of 16 participants) and future work should therefore have a closer look at such gender-related behavioral differences.


In addition, due to the high time demand for annotating the videos, six people were involved in the annotation process. Through training, calculation of the inter-coder reliability and repeated discussions we were trying to make the annotations as consistent as possible. Still, the annotations depend on subjective decisions by the annotators, because certain actions can not always be unambiguously assigned to one of the categories. However, as all videos of one participant were only coded by one annotator, we maximized the coding reliability within each participant as far as possible.

\section{Conclusions}
\deletemarker{We have reported findings obtained through reviewing and coding 559 hours of video material from a study in which participants were working in VR for an entire workweek.
This resulted in unique insights into how users behave when using VR HMDs for longer periods of time.}
\newmarker{We presented unique insights into how users behave during prolonged use of VR HMDs, obtained through reviewing and coding 559 hours of video material from a study in which participants worked in VR for an entire workweek.}
We found indications that participants are getting used to the HMD during the week, for example, the frequency of adjusting and supporting actions decreased over the five days of working in VR.
Also, participants seemed to be able to adapt to certain restrictions, as they \newmarker{mostly }did not take off the HMD\deletemarker{ for most of the time} while talking or using a phone.
On the other hand, by the end of the week participants removed the HMDs for longer periods of time\deletemarker{ and therefore also decreased the screen time}\newmarker{ which also resulted in less screen time}.
In addition, we found that the HMD can be disruptive to normal patterns of eating and drinking, as well as interacting with objects in the physical environment, as they were less frequent in VR.
These insights can be used to inform the design of less restricted, more ergonomic VR systems as tools for knowledge workers.
Overall, this paper presents detailed insights into user behavior while working in VR, which can only be detected through in-depth video analysis of data gathered over an extended period of time.
Therefore, we hope this work inspires further research on the long-term use of VR in various contexts.

\bibliographystyle{abbrv-doi-hyperref}
\balance
\bibliography{template}

\end{document}